\documentclass[amstex,twocolumn,showpacs,floats,floatfix,aps,pra]{revtex4}
\usepackage{amssymb}
\usepackage{amsmath}
\usepackage{calc}
\usepackage{graphicx}
\usepackage{bm}

\def\ket#1{\vert #1 \rangle}
\def\U{\mathbf{U}}
\def\fromto{\leftrightarrow}

\def\sec#1{{\bf #1}}

\begin{document}

\author{Andon A. Rangelov}
\affiliation{Department of Physics, Sofia University, James Bourchier 5 blvd., 1164 Sofia, Bulgaria}
\author{Nikolay V. Vitanov}
\affiliation{Department of Physics, Sofia University, James Bourchier 5 blvd., 1164 Sofia, Bulgaria}
\title{Complete population transfer in a three-state quantum system by a train of pairs of coincident pulses}
%\title{Discrete STIRAP}
\date{\today}

\begin{abstract}
A technique for complete population transfer between the two end states $\ket{1}$ and $\ket{3}$ of a three-state quantum system with a train of $N$ pairs of resonant and coincident pump and Stokes pulses is introduced.
A simple analytic formula is derived for the ratios of the pulse amplitudes in each pair for which the maximum transient population $P_2(t)$ of the middle state $\ket{2}$ is minimized, $P_2^{\max}=\sin^2(\pi/4N)$.
It is remarkable that, even though the pulses are on exact resonance, $P_2(t)$ is damped to negligibly small values even for a small number of pulse pairs.
The population dynamics resembles generalized $\pi$-pulses for small $N$ and stimulated Raman adiabatic passage for large $N$ and therefore this technique can be viewed as a bridge between these well-known techniques.
\end{abstract}

\maketitle

%%%%%%%%%%%%%%%%%%%%%%%%%%%%%%%%%%%%%%%%%%%%%%%%%%%%%%%%%%%%%%%%%%%%%%%%%%%%%%%%%%%%%%%%%%%%%%%%%%%%%%%%%%%%%%%%%%%%%%%%%%%%%%%%%%%%%%%%%%%%%%%%%%%%
%%%%%%%%%%%%%%%%%%%%%%%%%%%%%%%%%%%%%%%%%%%%%%%%%%%%%%%%%%%%%%%%%%%%%%%%%%%%%%%%%%%%%%%%%%%%%%%%%%%%%%%%%%%%%%%%%%%%%%%%%%%%%%%%%%%%%%%%%%%%%%%%%%%%
%%%%%%%%%%%%%%%%%%%%%%%%%%%%%%%%%%%%%%%%%%%%%%%%%%%%%%%%%%%%%%%%%%%%%%%%%%%%%%%%%%%%%%%%%%%%%%%%%%%%%%%%%%%%%%%%%%%%%%%%%%%%%%%%%%%%%%%%%%%%%%%%%%%%
\sec{Introduction.}
Coherent excitation of a discrete quantum system by an external resonant field represents an important notion in quantum mechanics.
Resonant pulses of specific pulse areas are widely used in a variety of fields, including coherent atomic excitation \cite{Allen-Eberly,Shore}, nuclear magnetic resonance \cite{Slichter}, quantum information \cite{NC}, and others.
Resonant excitation allows one to establish a complete control over the quantum system, particularly in two- and three-state systems.
Important examples include complete population transfer between the two states in a two-state system and between the two end states in a three-state $\Lambda$-system \cite{Shore,Vitanov2001a}.
Crucial conditions for resonant excitation are exact pulse areas and exact resonances between the frequencies of the external fields and the Bohr transition frequencies.
Deviations from exact resonances or exact pulse areas lead to deviations of the transition probability from the desired value.

To this end, an alternative to resonant excitation is provided by adiabatic passage techniques, which are robust against such deviations.
In three-state $\Lambda$-systems, the famous technique of stimulated Raman adiabatic passage (STIRAP) \cite{Gaubatz,Vitanov2001a,Vitanov2001b}
 allows complete population transfer between the two end states $\ket{1}$ and $\ket{3}$ in the adiabatic limit without placing any transient population in the middle state $\ket{2}$, even though the two driving fields --- pump and Stokes --- can be on exact resonance with their respective transitions, $\ket{1}\fromto\ket{2}$ and $\ket{2}\fromto\ket{3}$.
The conditions for STIRAP are the two-photon resonance between states $\ket{1}$ and $\ket{3}$, the counterintuitive pulse order (Stokes before pump), and adiabatic evolution.
However, adiabatic evolution requires large pulse areas, typically over $10\pi$ for smooth pulse shapes, which may be hard to reach experimentally.
Strategies for optimization of STIRAP with minimal pulse areas have been developed \cite{Vasilev,Dridi},
 but these come on the expense of strict relations on the pulse shapes \cite{Vasilev,Dridi} and require specific time-dependent nonzero detunings \cite{Dridi}.

%The reason is that throughout the adiabatic evolution of the system the population remains trapped in an adiabatic dark state, which is a superposition of states $\left\vert 1\right\rangle $ and $\left\vert 3\right\rangle $ only and does not involve the intermediate state $\left\vert 2\right\rangle $.
%Such a dark state is formed by maintaining a two-photon resonance between $\left\vert 1\right\rangle $ and $\left\vert 3\right\rangle $ during the interaction.
%If the pulses are ordered counterintuitively, the Stokes before the pump, then the dark state is associated with state $\left\vert 1\right\rangle $ initially and state $\left\vert 3\right\rangle $ in the end; thus providing an adiabatic route from $\left\vert 1\right\rangle $ to $\left\vert 3\right\rangle $.

In the present paper we introduce a novel technique that is both an alternative to the above techniques and reduces to them in two opposite limits.
The technique enables complete population transfer between the two end states $\ket{1}$ and $\ket{3}$ of a three-state $\Lambda$-system with a train of $N$ pairs of resonant and coincident pump and Stokes pulses with negligibly small transient population in the middle state $\ket{2}$, which vanishes as $N^{-2}$ with the number of pulse pairs $N$.
We note that unlike the adiabatic solutions, which are approximate, our technique is described by an exact analytic solution.
This technique formally resembles the techniques of piecewise adiabatic passage \cite{PAP-STIRAP,PAP-chirp} and composite pulses \cite{composite,composite-3}; the differences will be discussed toward the end.

We shall first present the exact analytic solution to the three-state dynamics for a single pulse pair and then the exact solution for a train of $N$ pairs of pulses,
 in which we shall demonstrate explicitly the dynamical suppression of the middle-state population.

%%%%%%%%%%%%%%%%%%%%%%%%%%%%%%%%%%%%%%%%%%%%%%%%%%%%%%%%%%%%%%%%%%%%%%%%%%%%%%%%%%%%%%%%%%%%%%%%%%%%%%%%%%%%%%%%%%%%%%%%%%%%%%%%%%%%%%%%%%%%%%%%%%%%
%%%%%%%%%%%%%%%%%%%%%%%%%%%%%%%%%%%%%%%%%%%%%%%%%%%%%%%%%%%%%%%%%%%%%%%%%%%%%%%%%%%%%%%%%%%%%%%%%%%%%%%%%%%%%%%%%%%%%%%%%%%%%%%%%%%%%%%%%%%%%%%%%%%%
%%%%%%%%%%%%%%%%%%%%%%%%%%%%%%%%%%%%%%%%%%%%%%%%%%%%%%%%%%%%%%%%%%%%%%%%%%%%%%%%%%%%%%%%%%%%%%%%%%%%%%%%%%%%%%%%%%%%%%%%%%%%%%%%%%%%%%%%%%%%%%%%%%%%
\sec{Single pulse pair.}
The probability amplitudes $c_k(t)$ ($k=1,2,3$) of the three states that form the $\Lambda$-system obey the Schr\"{o}dinger equation,
\begin{equation}\label{SEq}
i\hbar\partial_t\mathbf{c}(t) = \mathbf{H}(t) \mathbf{c}(t),
\end{equation}
where $\mathbf{c}(t)=[c_1(t),c_2(t),c_3(t)]^T$.
The Hamiltonian in the rotating-wave approximation \cite{Allen-Eberly,Shore} reads % (in units $\hbar =1$)%
\begin{equation}\label{H}
\mathbf{H}(t) =\frac{\hbar}{2}\left[\begin{array}{ccc}
0 & \Omega _{p}(t) & 0 \\
\Omega _{p}(t) & -i\Gamma & \Omega _{s}(t) \\
0 & \Omega _{s}(t) & 0
\end{array}\right],
\end{equation}
where $\Omega _{p}(t)$ and $\Omega _{s}(t)$ are the Rabi frequencies of the pump and Stokes pulses, respectively;
 each of them is proportional to the electric-field amplitude of the respective laser field and the corresponding transition dipole moment, $\Omega _{p}(t)=-\textbf{d}_{12}\cdot \textbf{E}_{p}(t)$ and $\Omega _{s}(t)=-\textbf{d}_{32}\cdot \textbf{E}_{s}(t)$.
 $\Gamma$ is the rate of irreversible loss from middle state $\ket{2}$.
For simplicity both $\Omega _{p}(t)$ and $\Omega _{s}(t)$ will be assumed real and positive because their phases can be eliminated by redefinition of the probability amplitudes. % as the populations do not depend on their phases.
More importantly, we assume that the Rabi frequencies are pulse-shaped functions with the same time dependence $f(t)$, but possibly with different magnitudes,
\begin{equation}
\Omega _{p,s}(t) = \Omega_{p,s}^0 f(t).  \label{the same time dependence}
\end{equation}
In this case --- single- and two-photon resonances and the same time dependence of the pump and Stokes fields ---
 the Schr\"odinger equation \eqref{SEq} is solved exactly by making a transformation to the so-called bright-dark basis \cite{Vitanov1998}.
The exact propagator reads \cite{Vitanov1998}
\begin{widetext}
\begin{equation}
\U(\theta) = \left[\begin{array}{ccc}
1- 2\sin ^{2}\theta\sin^2\frac14A  & -i \sin \theta\sin \frac12A  & - \sin 2\theta\sin^2\frac14A  \\
-i \sin \theta\sin \frac12A  & \cos \frac12A  & -i \cos \theta\sin \frac12A  \\
- \sin 2\theta\sin^2\frac14A  & -i \cos \theta\sin \frac12A  & 1-2 \cos^2\theta\sin^2\frac14A
\end{array}\right] ,  \label{evolution matrix}
\end{equation}
\end{widetext}
where the root-mean-square (rms) pulse area $A$ and the mixing angle $\theta $ are defined as
\begin{align}
A &= \int_{t_{i}}^{t_{f}}\sqrt{\Omega _{p}^{2}(t)+\Omega_{s}^{2}(t)}\, dt, \\
\tan \theta &= \frac{\Omega _{p}(t)}{\Omega _{s}(t)} = \frac{A_{p}}{A_{s}},
\end{align}
with $A_{p,s} = \int_{t_{i}}^{t_{f}} \Omega _{p,s}(t)\, dt$ being the pump and Stokes pulse areas.
 (For reasons that will become clear below we have omitted the area $A$ from the arguments of $\U$.)
Due to condition \eqref{the same time dependence} the angle $\theta$ is constant.

\begin{figure*}[t]
\includegraphics[width=120mm,angle=-90]{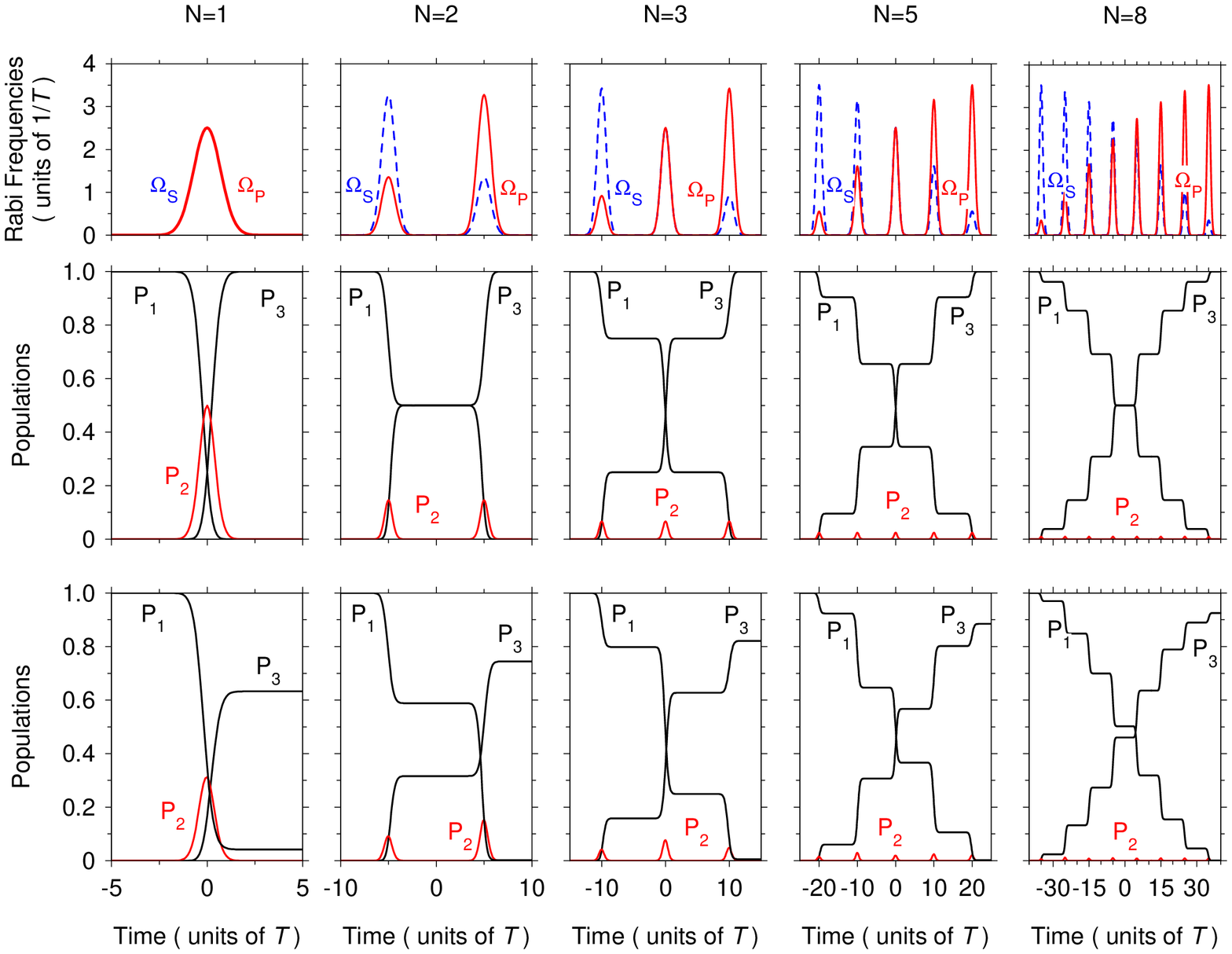}
\caption{
Rabi frequencies (top frames) and populations (bottom frames) vs time for (from left to right) 1, 2, 3, 5 and 8 pairs of pulses.
The pulse shapes are Gaussian, $\Omega_p(t)=\Omega\sin\theta_k e^{-(t-\tau_k)^2/T^2}$ and $\Omega_s(t)=\Omega\cos\theta_k e^{-(t-\tau_k)^2/T^2}$, with $\Omega=2\sqrt{\pi}/T$ (corresponding to rms pulse area $A=2\pi$)
 and the mixing angles $\theta_k$ ($k=1,2,\ldots,N$) are given by Eq.~\eqref{mixing angles}.
\emph{Middle frames:} no decay ($\Gamma=0$). Complete (stepwise) population transfer $1\to 3$ is achieved in all cases; however, the population $P_2(t)$ of the intermediate state is different.
The maximum of $P_2(t)$ is given by Eq.~\eqref{P2}; it vanishes as $1/N^2$.
\emph{Bottom frames:} irreversible loss from state $\ket{2}$ with a rate $\Gamma=1/T$.
}
\label{Fig1}
\end{figure*}

The propagator \eqref{evolution matrix} allows us to find the exact analytic solution for any initial condition;
 however, we restrict our attention here to a system initially in state $\ket{1}$: $\mathbf{c}(t_i)=(1,0,0)^T$.
Then the populations at the end of the interaction are
\begin{subequations}
\begin{align}
P_1(t_f) &= \left[1-2 \sin ^{2}\theta\sin^2\frac A4\right]^2,\\
P_2(t_f) &=  \sin^2 \theta \sin^2 \frac A2,\\
P_3(t_f) &=  \sin^2 2\theta \sin^4\frac A4.
\end{align}
\end{subequations}
Obviously, complete population transfer to state $\ket{3}$ is achieved for $\theta = \pi/4$, which corresponds to $A_p=A_s$, and rms pulse area $A=2\pi$.
Then, however, the intermediate state $\ket{2}$ acquires a transient population which reaches a maximum value of $P_2^{\max}=\frac12$
 at the intermediate time when the accumulated rms pulse area $A(t)$ reaches the mid-point value $\pi$.
This scenario is illustrated in Fig.~\ref{Fig1} (left frames).

We shall show below that the application of a train of resonant pulse pairs allows one to transfer the population from state $\ket{1}$ to state $\ket{3}$ completely,
 while reducing the transient population in state $\ket{2}$ to arbitrarily small value.

%\begin{widetext}
%-------------------------------------------------------------
%-------------------------------------------------------------
%\end{widetext}

%%%%%%%%%%%%%%%%%%%%%%%%%%%%%%%%%%%%%%%%%%%%%%%%%%%%%%%%%%%%%%%%%%%%%%%%%%%%%%%%%%%%%%%%%%%%%%%%%%%%%%%%%%%%%%%%%%%%%%%%%%%%%%%%%%%%%%%%%%%%%%%%%%%%
%%%%%%%%%%%%%%%%%%%%%%%%%%%%%%%%%%%%%%%%%%%%%%%%%%%%%%%%%%%%%%%%%%%%%%%%%%%%%%%%%%%%%%%%%%%%%%%%%%%%%%%%%%%%%%%%%%%%%%%%%%%%%%%%%%%%%%%%%%%%%%%%%%%%
%%%%%%%%%%%%%%%%%%%%%%%%%%%%%%%%%%%%%%%%%%%%%%%%%%%%%%%%%%%%%%%%%%%%%%%%%%%%%%%%%%%%%%%%%%%%%%%%%%%%%%%%%%%%%%%%%%%%%%%%%%%%%%%%%%%%%%%%%%%%%%%%%%%%
\sec{Train of pulse pairs.}
A sequence of $N$ pairs of pulses, each with rms pulse area $A$ and mixing angles $\theta_{k}$, produces the following total evolution matrix
\begin{equation}
\U^{(N)} = \U\left(\theta _{N}\right) \U\left(\theta_{N-1}\right) \cdots \U\left(\theta _{k}\right) \cdots \U\left(\theta_{1}\right) .  \label{overall evolution matrix}
\end{equation}
When the system is initially in state $\ket{1}$, the final populations are given by $P_n = \vert U^{(N)}_{n1} \vert^2$, with $n=1,2,3$.
Our objective is to have $P_1=P_2=0$ and $P_3=1$ at the end of the pulse train, with as little transient population in the middle state $\ket{2}$ as possible.

It is convenient to have an ``anagram'' pulse train that is symmetric with respect to time reversal, i.e., with mixing angles
%\begin{equation}
$\theta_{N+1-k}=\pi/2-\theta_k$, ($k=1,2,3...\lfloor N/2 \rfloor$).
%\end{equation}
In order to determine the values of $\theta _{k}$ and $A$, we use Eqs.~\eqref{evolution matrix} and \eqref{overall evolution matrix} and
 we first demand that after each pulse pair the population of state $\ket{2}$ vanishes;
 this gives immediately rms pulse area $A=2\pi$ for each pulse pair, hence the omission of $A$ from the arguments of $\U$ in Eq.~\eqref{evolution matrix}.

Next we require that the population is transferred to state $\ket{3}$ in the end, $P_1=P_2=0$ and $P_3=1$.
We further require that the maximum of the transient population $P_2(t)$ excited by each pulse pair is the same.
Among the many solutions we pick the one that minimizes the maxima of $P_2(t)$.
A simple algebra gives the angles
\begin{equation}\label{mixing angles}
\theta _{k}=\frac{(2k-1) \pi }{4N} \quad  (k=1,2,3...N).
\end{equation}

The maximum population of state $\ket{2}$, which, as for a single pulse pair above, occurs in the middle of each pulse pair (at rms area $\pi$), is readily obtained,
\begin{equation}
P_{2}^{\max} %=\left\vert C_{2}(t)\right\vert _{\max }^{2}
 =\sin ^{2}\left( \frac{\pi}{4N}\right) . \label{P2}
\end{equation}
From here we conclude immediately that for large $N$ the maximum population of the middle state vanishes as $1/N^2$.
Obviously, for $N\geqq 8$ pairs, the transient population in state $\ket{2}$ does not exceed 1\%.
It is particularly significant that this suppression occurs on resonance and it results from the destructive interference of the successive interaction steps, rather than from a large detuning.
We note that for $N\gg 1$ the total pulse area is very large, which is the condition for adiabatic evolution on resonance in STIRAP.

We show in Fig.~\ref{Fig1} the population evolution for several pulse trains of different number of pulse pairs $N$.
In all cases the population is transferred from state $\ket{1}$ to state $\ket{3}$ in the end in a stepwise manner.
The transient population of the intermediate state $\ket{2}$ is damped as $N$ increases: from 0.5 for a single pair of pulses and 0.15 for two pairs to below 1\% for 8 pulse pairs.
For small $N$ the transition picture resembles (fractional) resonant excitation, whereas for large $N$ the transition picture resembles adiabatic passage.
The lower frames demonstrate the effect of irreversible population loss from state $\ket{2}$ for a loss rate $\Gamma=1/T$.
As it can be expected from the middle frames of lossless interaction,
 the damping of the intermediate-state population by longer pulse trains reduces the effect of population loss:
 the target-state population $P_3$ decreases to 0.63 for a single pulse pair but just to 0.93 for a train of $N=8$ pulse pairs.
This population loss can be decreased even further by longer trains.

We note that because the technique uses fields on exact resonance the pulse shapes are unimportant.
Although the example in Fig.~\ref{Fig1} uses Gaussian shapes, pulses of rectangular or any other shape are equally suitable.

%%%%%%%%%%%%%%%%%%%%%%%%%%%%%%%%%%%%%%%%%%%%%%%%%%%%%%%%%%%%%%%%%%%%%%%%%%%%%%%%%%%%%%%%%%%%%%%%%%%%%%%%%%%%%%%%%%%%%%%%%%%%%%%%%%%%%%%%%%%%%%%%%%%%
%%%%%%%%%%%%%%%%%%%%%%%%%%%%%%%%%%%%%%%%%%%%%%%%%%%%%%%%%%%%%%%%%%%%%%%%%%%%%%%%%%%%%%%%%%%%%%%%%%%%%%%%%%%%%%%%%%%%%%%%%%%%%%%%%%%%%%%%%%%%%%%%%%%%
%%%%%%%%%%%%%%%%%%%%%%%%%%%%%%%%%%%%%%%%%%%%%%%%%%%%%%%%%%%%%%%%%%%%%%%%%%%%%%%%%%%%%%%%%%%%%%%%%%%%%%%%%%%%%%%%%%%%%%%%%%%%%%%%%%%%%%%%%%%%%%%%%%%%
\sec{Discussion.}
We have demonstrated that complete population transfer in a three-state system driven by a pair of external pulse-shaped laser fields that are on resonance and have the same time dependence,
 can be accomplished in such a way that all population is transferred from the initial state $\ket{1}$ to the target state $\ket{3}$
  with minimal transient population in the middle state $\ket{2}$.
This is achieved with a train of $N$ pairs of coincident pulses with appropriately chosen amplitudes.
In the limit $N\gg 1$, the present technique resembles STIRAP, because then the successive increments of the mixing angle $\theta_k$ become very small (nearly continuous),
 with the Stokes field dominating over the pump field in the beginning and then the pump field dominating in the end,
 in exact analogy to the counterintuitive sequence Stokes-pump in STIRAP \cite{Gaubatz,Vitanov2001a,Vitanov2001b}.
However, the present technique achieves complete population transfer $1\to3$ also for small $N$, in a manner reminiscent of generalized $\pi$-pulses \cite{Shore}, while keeping the middle-state population at very low values.
In this manner, the present technique can be viewed as a bridge between generalized $\pi$-pulses and STIRAP.
%, hence we can name it ``discrete STIRAP''.

The present technique can be viewed also as an alternative of two other techniques that use pulse trains for complete population transfer.
Piecewise adiabatic passage (PAP) \cite{PAP-STIRAP,PAP-chirp} uses a train of a large number of pulses, each of which produces a perturbatively small change in the populations,
 while the present technique works for an arbitrary number of pulse pairs and each pair may produce a large population change (for small $N$).
In one of the implementations involving only two states \cite{PAP-chirp}, PAP demands phases that change quadratically from pulse to pulse, which translate into a linear chirp for a large number of pulses;
 the present technique does not need such quadratic phases but appropriate amplitude ratios.
% the population evolution is a piecewise version of the one for standard single-pulse AP.
In the PAP implementation with three states \cite{PAP-STIRAP}, to which the present technique is more closely related, the pump and Stokes fields in STIRAP are turned abruptly on and off repeatedly;
 the amplitudes of the individual pulses are determined such that they match the envelopes of the pump and Stokes pulses in STIRAP.
In the present technique the pulse amplitudes are determined from the conditions to achieve complete population transfer to the target state $\ket{3}$ and to minimize the population of the middle state $\ket{2}$.
Indeed, the systematic suppression of the transient population of state $\ket{2}$ with the increasing number of pulse pairs $N$ seen in Fig.~\ref{Fig1} is only observed in the present technique.
We also point out that the solution in the present paper is exact while PAP and STIRAP give approximate solutions in the adiabatic limit.

The present technique is also reminiscent of the technique of composite pulse sequences \cite{composite}.
The latter uses sequences of pulses in two-state systems \cite{composite}, or sequences of pulse pairs in three-state systems \cite{composite-3}, with well-defined relative phases, which are determined from the condition to produce a desired excitation profile.
Therefore the control parameters in the composite pulses are the relative phases, whereas in the present technique the control parameters are the amplitude ratios in each pulse pair.
Moreover, the objective in the composite pulse technique is the shape of the excitation profile while in our technique the main objective, beside the complete population transfer $1\to 3$, is the suppression of the intermediate-state population.

%%%%%%%%%%%%%%%%%%%%%%%%%%%%%%%%%%%%%%%%%%%%%%%%%%%%%%%%%%%%%%%%%%%%%%%%%%%%%%%%%%%%%%%%%%%%%%%%%%%%%%%%%%%%%%%%%%%%%%%%%%%%%%%%%%%%%%%%%%%%%%%%%%%%
%%%%%%%%%%%%%%%%%%%%%%%%%%%%%%%%%%%%%%%%%%%%%%%%%%%%%%%%%%%%%%%%%%%%%%%%%%%%%%%%%%%%%%%%%%%%%%%%%%%%%%%%%%%%%%%%%%%%%%%%%%%%%%%%%%%%%%%%%%%%%%%%%%%%
%%%%%%%%%%%%%%%%%%%%%%%%%%%%%%%%%%%%%%%%%%%%%%%%%%%%%%%%%%%%%%%%%%%%%%%%%%%%%%%%%%%%%%%%%%%%%%%%%%%%%%%%%%%%%%%%%%%%%%%%%%%%%%%%%%%%%%%%%%%%%%%%%%%%
\sec{Conclusion.}
The technique introduced in this paper allows complete population transfer between states $\ket{1}$ and $\ket{3}$ via an intermediate state $\ket{2}$ with a train of $N$ pairs of coincident pump and Stokes pulses, by placing only a negligible transient population in state $\ket{2}$, which decreases as $1/N^2$ as the number of pulse pairs $N$ increases.
This technique resembles the technique of generalized $\pi$-pulses for small $N$ and STIRAP for large $N$ and therefore it can be viewed as a bridge between these two well-known techniques.
It is remarkable that the middle-state population $P_2(t)$ is damped considerably even for a small number of pulse pairs despite the fact that the pump and Stokes fields are on exact resonance with their transitions.
All these features make this technique an interesting alternative of the existing techniques for coherent control of three-state quantum systems.

%%%%%%%%%%%%%%%%%%%%%%%%%%%%%%%%%%%%%%%%%%%%%%%%%%%%%%%%%%%%%%%%%%%%%%%%%%%%%%%%%%%%%%%%%%%%%%%%%%%%%%%%%%%%%%%%%%%%%%%%%%%%%%%%%%%%%%%%%%%%%%%%%%%%
This work is supported by the European network FASTQUAST and the Bulgarian NSF grants D002-90/08, DMU02-19/09 and IRC-COSIM.

%%%%%%%%%%%%%%%%%%%%%%%%%%%%%%%%%%%%%%%%%%%%%%%%%%%%%%%%%%%%%%%%%%%%%%%%%%%%%%%%%%%%%%%%%%%%%%%%%%%%%%%%%%%%%%%%%%%%%%%%%%%%%%%%%%%%%%%%%%%%%%%%%%%%
%%%%%%%%%%%%%%%%%%%%%%%%%%%%%%%%%%%%%%%%%%%%%%%%%%%%%%%%%%%%%%%%%%%%%%%%%%%%%%%%%%%%%%%%%%%%%%%%%%%%%%%%%%%%%%%%%%%%%%%%%%%%%%%%%%%%%%%%%%%%%%%%%%%%
%%%%%%%%%%%%%%%%%%%%%%%%%%%%%%%%%%%%%%%%%%%%%%%%%%%%%%%%%%%%%%%%%%%%%%%%%%%%%%%%%%%%%%%%%%%%%%%%%%%%%%%%%%%%%%%%%%%%%%%%%%%%%%%%%%%%%%%%%%%%%%%%%%%%

\end{document}